\documentstyle[12pt,psfig]{article}

\newcommand{\resection}[1]{\setcounter{equation}{0}\section{#1}}

\newcommand{\EQ}{\begin{equation}}
\newcommand{\EN}{\end{equation}}
\newcommand{\bea}{\begin{eqnarray}}
\newcommand{\eea}{\end{eqnarray}}

\newcommand{\th}{\theta}
\newcommand{\var}{\varepsilon}

\begin{document}
\topmargin 0pt
\oddsidemargin 5mm
\renewcommand{\thefootnote}{\arabic{footnote}}
\newpage
\begin{flushright}
LPTHE/99-08
\end{flushright}
\vspace{0.5cm}
\begin{center}
{\large {\bf First order phase transitions and integrable field theory.

The dilute $q$-state Potts model}}\\
\vspace{1.8cm}
{\large G. Delfino} \\ \vspace{0.5cm}
{\em Laboratoire de Physique Th\'eorique et Hautes Energies}\\
{\em Universit\'e Pierre et Marie Curie, Tour 16 $1^{er}$ \'etage, 4 place 
Jussieu}\\
{\em 75252 Paris cedex 05, France}\\
{\em E-mail: aldo@lpthe.jussieu.fr}\\
\end{center}
\vspace{1.2cm}

\renewcommand{\thefootnote}{\arabic{footnote}}
\setcounter{footnote}{0}

\begin{abstract}
\noindent
We consider the two-dimensional dilute $q$-state Potts model on its first 
order phase transition surface for $0<q\leq 4$. After determining the exact
scattering theory which describes the scaling limit, we compute the two-kink
form factors of the dilution, thermal and spin operators. They provide an 
approximation for the correlation functions whose accuracy is 
illustrated by evaluating the central charge and the scaling dimensions along
the tricritical line.
\end{abstract}

\vspace{.3cm}


\newpage
\resection{Introduction}

Two-dimensional quantum field theory allows for a large class of 
non-scale-invariant models which exhibit an infinite number of integrals of 
motion \cite{Taniguchi}. These models are {\em integrable}, in the sense 
that the $S$-matrix which determine their on-shell physics in $1+1$ dimensions
can be computed exactly \cite{ZZ}. The solvability of the models extends
off-shell through the possibility of exact determination of the matrix elements
of local operators on the asymptotic states \cite{Karowski,Smirnov}, which
in turn leads to the evaluation of correlation functions in the form of 
spectral series. 

A natural domain of application for quantum field theory is provided by 
the scaling limit of classical statistical mechanics. It is well known that in
two dimensions there exists a number of lattice models which are exactly 
solvable due to the presence of a family of commuting transfer matrices 
\cite{Baxter}. The scaling limit of these models is 
described by an integrable field theory. What is more interesting is that
the opposite appears to be not true, a circumstance which attests of an 
inequivalence 
of the notions of integrability on the lattice and in the continous limit. 
In fact, there are integrable quantum field theories which describe the 
scaling limit of statistical models which are {\em not} solved on the 
lattice. For these models, integrability seems to emerge as a property of the
scaling limit, after that the non-universal, lattice dependent features have 
been eliminated. Among the examples of this kind which have been studied in 
more detail, we mention the Ising model in a magnetic field at $T=T_c$ 
\cite{Taniguchi,immf,DS}, the off-critical $q$-state Potts model with 
$0<q\leq 4$ \cite{CZ,DC}, and the off-critical Ashkin-Teller model \cite{AT}.

In this paper we are interested in the generalisation of the $q$-state Potts 
model which allows for the presence of annealed vacancies on the lattice 
\cite{Wu}.
More precisely, we will consider the model on its first order transition
surface for $q\leq 4$. The model is not solvable on the lattice but is 
integrable in the scaling limit because it corresponds to the perturbation
of the conformal field theories with $C\leq 1$ by the primary operator 
$\varphi_{1,3}$ \cite{DF}. The $\varphi_{1,3}$-perturbed conformal field 
theories have done the object of extensive study. In particular, they are 
known to describe the scaling limit of the RSOS models \cite{ABF,Huse} and
the $O(n)$ model for $-2<n<2$ \cite{Nienhuis,DF}. The exact scattering 
descriptions 
corresponding to the massive phases of these models were found in 
Refs.\,\cite{SashaRSOS,LeClair} and \cite{Selfavoiding}, respectively. 
Here we will directly determine the $S$-matrix which is suitable for the
Potts model interpretation by requiring that it describes a system in which
a disordered phase coexists with $q$ ordered phases which are exchanged by
permutation symmetry. The cases $q=2,3$ were already discussed in 
Ref.\,\cite{fractional}. 

Working within the particle basis which makes 
explicit the relevant symmetries of the model is important for dealing with
the problem of the determination of the matrix elements of local operators
on the asymptotic particle states (form factors).
The knowledge of the form factors determines the correlation functions in
the form of spectral series. These series, however, are non-trivial 
mathematical objects for which no resummation technique is presently available.
For this reason, it becomes important to be able to judge of the level of 
accuracy yielded by partial sums. A particularly interesting test is provided
by the sum rules which express the ultravioled data (central charge and 
scaling dimensions) as integrals of correlation functions computed in the 
off-critical theory \cite{cth,DSC}. The general response is that the rate
of convergence of the spectral series is remarkably fast. The results of this
paper provide, in particular, an illustration of this fact. Of course, the 
agreement with the predictions of conformal field theory for the ultraviolet
limit also represents a non-trivial confirmation of the correctness of the 
proposed scattering theory. 

The layout of the paper is the following. In the next section we recall some
basic features of the dilute Potts model before turning to the determination
of the $S$-matrix in section 3. The exact scattering description is then
used in section 4 to compute the form factors of the dilution, thermal and 
spin operators. Section 5 deals with the correlation functions of these 
operators and the sum rules for the central charge and the scaling dimensions;
correlation lengths and interfacial tensions are also discussed.
Few final remarks are collected in section 6.

\resection{The dilute Potts model}

The dilute $q$-state Potts model is defined by the lattice Hamiltonian
\cite{Wu}
\EQ
H=-J\sum_{\langle x,y\rangle}t(x)t(y)\delta_{s(x),s(y)}+\Delta\sum_xt(x)\,\,.
\label{lattice}
\EN
Here $s(x)$ is the ordinary Potts spin variable which can assume $q$ different 
values (colours); the system is clearly invariant under permutations of the 
colours. $t(x)$ is a lattice gas variable which equals $0$ if the site $x$ is 
empty or $1$ if it is occupied. The coupling $J>0$ (ferromagnetic case) 
plays the role of the inverse temperature, while $\Delta$ is a chemical 
potential controlling the vacancy density. The pure Potts model is recovered
when $\Delta=-\infty$. 

When writing the Hamiltonian (\ref{lattice}) it is natural to think of the
number of colours $q$ as being an integer. However, it is well known that
it makes sense and is extremely useful to consider the analyitically 
continued version of the Potts model in which $q$ is a real parameter.
In the pure model, the ``monochromatic'' limit $q\rightarrow 1$ describes
isotropic percolation \cite{KF}. The dilute system simply reduces to an 
Ising model at $q=1$.

The phase diagram corresponding to the Hamiltonian (\ref{lattice}) in two 
dimensions \cite{NBRS} is shown in Fig.\,1. The system undergoes a 
ferromagnetic phase 
transition at a critical value of the temperature $T_c$. It is desordered
at $T>T_c$, and exhibits spontaneous magnetisation in the low-temperature 
phase in which $q$ vacua (one for each colour) are degenerate.
In the pure model ($e^\Delta=0$), the phase transition is second order for 
$q\leq 4$ and first order for $q>4$ \cite{Baxter}. 
When the vacancy density becomes non-zero, the transition remains 
continous at $q<4$ up to a critical value $\Delta_c$ of the dilution, and
becomes first order for $\Delta>\Delta_c$. The disordered phase
coexists with the $q$ ordered phases along the first order transition lines
(dashed in Fig.\,1). The line $T=T_c$, $\Delta=\Delta_c$ (upper thick segment 
in the phase diagram) is a line of tricritical fixed points. Since the 
transition is 
always discontinous for $q>4$, the tricritical line has to meet the line of 
fixed points of the pure model (lower thick segment) at $q=4$.

Let us consider the dilute $q$-state Potts model in the framework of 
conformal field theory and its perturbations. Both the critical and  
tricritical line of fixed point are described by conformal field theory
\cite{DF}. Here we are interested in the tricritical one, since it describes
the ultraviolet limit of the first order phase transition lines we want to
study. It corresponds to the conformal theory with central charge
\EQ
C=1-\frac{6}{p(p+1)}\,,\hspace{1cm}p>2
\label{c}
\EN
with $p$ related to $q$ as
\EQ
\sqrt{q}=2\cos\frac{\pi}{p}\,\,.
\label{pq}
\EN
In the continous field theoretic description, the dilution, thermal and
spin variables are described by operators which we denote by $\psi(x)$, 
$\var(x)$ and $\sigma_i(x)$ ($i=1,2,\ldots,q$), respectively. They correspond
to the operators $\varphi_{1,3}$, $\varphi_{1,2}$ and $\varphi_{\frac{p}{2},
\frac{p}{2}}$ in the conformal field theory classification, and their 
scaling dimensions along the tricritical line are
\bea
&& X_\psi=2\,\frac{p-1}{p+1}\,\,,\\
&& X_\var=\frac{p-2}{2\,(p+1)}\,\,,\label{x12}\\
&& X_\sigma=\frac{p^2-4}{8p\,(p+1)}\,\,.\label{xsig}
\eea
The renormalisation group trajectories flowing out of the tricritical line
at $T=T_c$ (vertical lines in Fig.\,1) are described by the perturbation of 
the tricritical line by the dilution operator $\psi(x)$, which is relevant 
($X_\psi<2$) for $q<4$. The action associated to such trajectories 
reads
\EQ
{\cal A}={\cal A}_{\mbox{tricr}}+g\int d^2x\,\psi(x)\,,
\label{action}
\EN
where ${\cal A}_{\mbox{tricr}}$ denotes the action of the tricritical line
and $g\sim\Delta-\Delta_c$ is the coupling measuring the deviation from the 
critical dilution. Depending on the sign of $g$, the action above describes
either the massless flows to the ordinary critical point (at least for 
$2\leq q<4$), or the massive trajectories associated to the first order 
transition. Both types of trajectories are known to be 
integrable, as a consequence of the integrability of the $\varphi_{1,3}$ 
perturbations of conformal field theory \cite{Taniguchi}. 

In this paper, we deal with the first order trajectories along which the 
elementary excitations have a mass $m$ (inverse correlation length) related
to the coupling $g$ as
\EQ
m\sim g^\frac{1}{2-X_\psi}\,\,,\hspace{1cm}q<4\,\,.
\EN
As $q$ approaches $4$ from below, the perturbation becomes marginally relevant,
the correlation length develops an essential singularity as $g\rightarrow 0^+$,
and the previous relation is replaced by \cite{NSC}
\EQ
m\sim e^{-A/g}\,\,,\hspace{1cm}q=4
\label{mg4}
\EN
where $A$ is a positive constant. Notice that for $q=1$ one has $C=1/2$, 
$X_\psi=1$ and $X_\var=1/8$, so that the action (\ref{action}) describes
an Ising model in zero magnetic field.

\resection{Scattering theory}

In this section we determine the exact scattering theory describing the 
first order transition lines of the dilute $q$-state Potts model (vertical
dashed lines in Fig.\,1). In 1+1 dimensions one knows that the elementary
excitations of a theory exhibiting degenerate ferromagnetic vacua are kinks 
interpolating between adjacent vacua. Kink excitations differ from the 
ordinary particle excitations for the fact that, in general, the composition
of multi-kink states is subject to some restrictions. Both the number of 
elementary kinks and the type of restrictions can be argued from an analysis
of the vacuum structure.

Consider the dilute Potts model at $T\leq T_c$. There are $q$ ordered  
ground states that we denote $\Omega_i$, $i$ being the colour index; they are 
sent into one another by the permutation group under which the model is 
invariant. Intuitively, one can think of them as being located at the $q$ 
vertices of a hypertetrahedron living in the $(q-1)$-dimensional space 
of the independent spin components. The disordered vacuum (we call it 
$\Omega_0$) is located at the center of the hypertetrahedron.
At $T<T_c$ the energy of the disordered vacuum is higher than that of the 
ordered vacua and the elementary excitations of the 
scattering theory are kinks\footnote{The rapidity $\th$ parameterises the
on-shell energy and momentum of the kinks of mass $m$ as 
$(p^0,p^1)=(m\cosh\th,m\sinh\th)$.} $\tilde{K}_{ij}(\th)$ interpolating among 
the ordered vacua. They can be associated to the edges of the hypertetrahedron.
This scattering theory and its off-shell consequences were studied 
in Refs.\,\cite{CZ,DC} for the case of the pure Potts model. On the first 
order transition surface ($T=T_c$, $\Delta>\Delta_c$)
the disordered vacuum becomes degenerate with the ordered vacua and the 
$q+1$ phases coexist. In this situation the elementary excitations are 
the kinks $K_{0i}(\th)$ and $K_{i0}(\th)$ interpolating among 
the center of the hypertetrahedron and its vertices, while the kinks we had at 
$T<T_c$ become composite excitations\footnote{In principle, $K_{i0}K_{0i}$ 
could also form bound states corresponding to $\tilde{K}_{ij}$, but we 
will see that this is not the case.} -- $\tilde{K}_{ij}\sim K_{i0}K_{0j}$.

Let us fix our attention on the transition surface 
and consider the space of states constructed on the elementary excitations
$K_{i0}$, $K_{0i}$. The interpretation of kinks as 
excitations interpolating among adjacent vacua requires that, in a 
multi-kink state, adjacent vacuum indices belonging to different kinks 
coincide. In our case, this means that the only allowed kink sequences are of 
the type
\EQ
\ldots K_{0i}(\th_1)K_{i0}(\th_2)K_{0j}(\th_3)K_{j0}(\th_4)\ldots\,\,\,.
\EN 
The integrability of the theory ensures that the scattering of these 
multi-kink states is completely 
elastic and factorised into the product of elementary two-body amplitudes
\cite{ZZ}. Taking into account the vacuum structure described above and the 
permutation symmetry of the ordered vacua, one immediately realises that the 
scattering theory has only the four two-kink amplitudes represented in Fig.\,2.
They can be associated to the commutation relations
\bea
K_{0i}(\th_1)K_{i0}(\th_2)=A_0(\th_1-\th_2)K_{0i}(\th_2)K_{i0}(\th_1)+
                A_1(\th_1-\th_2)\sum_{j\neq i}K_{0j}(\th_2)K_{j0}(\th_1),
\label{fz1} \\
K_{i0}(\th_1)K_{0j}(\th_2)=
                   \delta_{ij}B_0(\th_1-\th_2)K_{i0}(\th_2)K_{0i}(\th_1)+
               (1-\delta_{ij})B_1(\th_1-\th_2)K_{i0}(\th_2)K_{0j}(\th_1).
\label{fz2}
\eea
The amplitudes are related in pairs by crossing symmetry
\bea
&& A_0(\th)=B_0(i\pi-\th)\,,\label{crossing}\\
&& A_1(\th)=B_1(i\pi-\th)\,\,.\nonumber
\eea
They also have to satisfy the unitarity conditions which can be formally 
obtained by commuting once again the r.h.s. of Eqs.\,(\ref{fz1}), (\ref{fz2})
\bea
&& A_0(\th)A_0(-\th)+(q-1)A_1(\th)A_1(-\th)=1\,,\label{u1}\\
&& A_0(\th)A_1(-\th)+(q-2)A_1(\th)A_1(-\th)+A_1(\th)A_0(-\th)=0\,,\label{u2}\\
&& B_0(\th)B_0(-\th)=B_1(\th)B_1(-\th)=1\,\,.\label{u3}
\eea
Finally, the so-called factorisation (or Yang-Baxter) equations are obtained
by considering a three-kink initial state and requiring the equality of the 
final states obtained performing the pair commutations (\ref{fz1}), (\ref{fz2})
in the two possible orderings
\bea
&& A_0B_0A_0+(q-1)A_1B_1A_1=B_0A_0B_0\,,\label{yb1}\\
&& A_0B_0A_1+A_1B_1A_0+(q-2)A_1B_1A_1=B_1A_1B_0\,,\\
&& A_0B_1A_0+A_1B_0A_1+(q-2)A_1B_1A_1=B_1A_0B_1\,,\\
&& A_0B_1A_1+A_1B_0A_1+A_1B_1A_0+(q-3)A_1B_1A_1=B_1A_1B_1\,;\label{yb3}
\eea
here the arguments of the three factors in each product are $\th$, $\th+\th'$
and $\th'$, respectively.

A solution to Eqs.\,(\ref{crossing})--(\ref{yb3}) can be found along the 
following lines.
It is easily seen that they fix $A_1(0)=0$, $A_0(0)=B_0(0)=B_1(0)=
\pm 1$. Then, remembering also crossing and unitarity, we know that 
$R\equiv B_1/B_0$ satisfies $R(\th)R(-\th)=1$, $R(0)=1$ and $R(i\pi)=0$. The
analysis of the cases $q=2,3$ in Ref.\,\cite{fractional} suggests the ansatz 
$R(\th)=\sinh\lambda(i\pi-\th)/\sinh\lambda(i\pi+\th)$. Then Eq.\,(\ref{u2})
immediately fixes $2\cos\pi\lambda=\sqrt{q}$ or, in view of (\ref{pq}), 
$\lambda=1/p$. Notice that the restriction $q\leq 4$ automatically emerges 
from the consistency requirements of the scattering theory. This is expected
since the scaling limit of the lattice model can no longer be defined
for $q>4$. 

Having determined the ratio $A_1(\th)/A_0(\th)=R(i\pi-\th)$, we can reduce 
Eq.\,(\ref{u1}) to the form
\EQ
A_1(\th)A_1(-\th)=-\frac{1}{4\cos^2\frac{\pi}{p}}\,
\frac{\sinh^2\frac{\th}{p}}{\sinh\frac{1}{p}(i\pi+\th)
                               \sinh\frac{1}{p}(i\pi-\th)}\,\,.
\nonumber
\EN
Solving this equation together with $A_1(i\pi-\th)A_1(i\pi+\th)=1$ leads to 
the final result
\bea
&& A_0(\th)=\frac{e^{-i\gamma\th}}{\sqrt{q}}\,
 \frac{\sinh\frac{1}{p}(2i\pi-\th)}{\sinh\frac{1}{p}(i\pi-\th)}\,S_0(\th)\,,
\label{a0}\\
&& A_1(\th)=\frac{e^{-i\gamma\th}}{\sqrt{q}}\,
 \frac{\sinh\frac{1}{p}\th}{\sinh\frac{1}{p}(i\pi-\th)}\,S_0(\th)\,,\\
&& B_0(\th)=e^{i\gamma\th}\,
 \frac{\sinh\frac{1}{p}(i\pi+\th)}{\sinh\frac{1}{p}(i\pi-\th)}\,S_0(\th)\,,\\
&& B_1(\th)=e^{i\gamma\th}\,S_0(\th)\,,
\label{b1}
\eea
with
\EQ
\gamma=\frac{1}{2\pi}\ln q\,\,,
\EN
\bea
S_0(\th)=-\prod_{n=0}^\infty\frac{
\Gamma\left(1+\frac{2}{p}(n+\frac{1}{2})+\frac{\th}{i\pi p}\right)
\Gamma\left(1+\frac{2}{p}n-\frac{\th}{i\pi p}\right)}{
\Gamma\left(1+\frac{2}{p}(n+\frac{1}{2})-\frac{\th}{i\pi p}\right)
\Gamma\left(1+\frac{2}{p}n+\frac{\th}{i\pi p}\right)}&&\nonumber \\
         \times\frac{
\Gamma\left(\frac{2}{p}(n+1)-\frac{\th}{i\pi p}\right)
\Gamma\left(\frac{2}{p}(n+\frac{1}{2})+\frac{\th}{i\pi p}\right)}{
\Gamma\left(\frac{2}{p}(n+1)+\frac{\th}{i\pi p}\right)
\Gamma\left(\frac{2}{p}(n+\frac{1}{2})-\frac{\th}{i\pi p}\right)}&&\nonumber \\
        =-\exp\left\{i\int_0^\infty\frac{dx}{x}\,
\frac{\sinh(p-1)\frac{x}{2}}{\sinh\frac{px}{2}\cosh\frac{x}{2}}\,
\sin\frac{x\th}{\pi}\right\}\,\,.\hspace{1cm}&&
\eea
The amplitudes above satisfy the factorisation equations 
(\ref{yb1})--(\ref{yb3}).
They are free of poles in the physical strip $\mbox{Im}\th
\in(0,\pi)$ and then the theory does not possess bound states. It can be 
checked that at $q=1$ the amplitudes $A_0$ and $B_0$ reduce to $-1$, as 
expected for the thermal Ising model\footnote{More precisely, we used this 
condition at
$q=1$ to fix the overall sign of the $S$-matrix which is left undetermined
by the general equations.}. At $q=4$ the amplitudes (\ref{a0})--(\ref{b1}) 
become linear combinations of those of Sine-Gordon solitons\footnote{For 
generic values of $p$, the function $S_0(\th)$ appears in the Sine-Gordon 
$S$-matrix. This analogy reflects the well known circumstance that 
$\varphi_{1,3}$-perturbed conformal field theories are related to suitable 
restrictions of the Sine-Gordon model \cite{Smirnov13,LeClair}.} 
at the point $\beta=8\pi$ where the term $\cos\beta\varphi$ which perturbs
the gaussian fixed point becomes marginally relevant. This scattering theory
is known to imply the relation (\ref{mg4}) among the mass and the coupling
constant \cite{AlioshaSG}. These
checks should be sufficient to to get rid of the doubts concerning the usual
CDD ambiguities\footnote{For $q=2,3$, the amplitudes (\ref{a0})--(\ref{b1}) 
do not completely coincide
with those proposed in Ref.\,\cite{fractional}. It can be checked that
the infinite products of gamma functions in \cite{fractional}, although 
formal solutions of all the equations, in fact converge to zero.}.

\resection{Form factors}

The $S$-matrix is not an object of primary interest for statistical 
mechanics. The link with the off-shell phyics is provided by the matrix 
elements of the local operators on the asymptotic particle states.
They are known as form factors and are exactly
computable in integrable quantum field theories \cite{Karowski,Smirnov}.
In this section we consider the matrix elements computed on two-kink states;
they are relatively easy to determine and prove sufficient to provide 
accurate quantitative information on the correlation functions.

We are interested in the operators $\Theta(x)\sim\psi(x)$, $\var(x)$ and
$\sigma_j(x)$, $\Theta(x)$ denoting the trace of the stress-energy tensor.
They all couple to states with zero topological charge, i.e. excitations 
which begin and end on the same vacuum. Depending on whether the latter is
the disordered vacuum $\Omega_0$ or one of the ordered vacua 
$\Omega_i$, we have the two-kink form factors (Fig.\,3)
\bea
&& \langle\Omega_0|\Phi(0)|K_{0i}(\th_1)K_{i0}(\th_2)\rangle\equiv 
F^\Phi_{0i}(\th_1-\th_2)\,,\label{ff1}\\
&& \langle\Omega_i|\Phi(0)|K_{i0}(\th_1)K_{0i}(\th_2)\rangle\equiv 
F^\Phi_{i0}(\th_1-\th_2)\label{ff2}\,,
\eea
where $\Phi$ generically denotes one of the above mentioned operators.
More specifically, the operators $\Theta$ and $\var$ are invariant under
permutations of the colours, and we can write
\EQ
F^\Phi_{0i}(\th)=F^\Phi_-(\th)\,,\hspace{1cm}\Phi=\Theta,\var
\EN
\EQ
F^\Phi_{i0}(\th)=F^\Phi_+(\th)\,,\hspace{1cm}\Phi=\Theta,\var\,\,.
\EN
The spin operators $\sigma_j(x)$ are related to the lattice variables $s(x)$
as 
\EQ
\sigma_j(x)=\delta_{s(x),j}-\frac{1}{q}\,,\hspace{1cm}j=1,2,\ldots,q
\EN 
the constant being subtracted to ensure the vanishing of the order parameter
$\langle\sigma_j\rangle$ in the disordered phase. When considering the matrix 
elements 
(\ref{ff1}), (\ref{ff2}) with $\Phi=\sigma_j$, we only need to distinguish 
whether the colour indices $i$ and $j$ are equal or different. Since 
$\sum_{j=1}^q\sigma_j=0$, we can write
\EQ
F^{\sigma_j}_{0i}(\th)=\frac{q\delta_{ij}-1}{q-1}\,F^\sigma_-(\th)\,,
\label{constraint}
\EN
\EQ
F^{\sigma_j}_{i0}(\th)=\frac{q\delta_{ij}-1}{q-1}\,F^\sigma_+(\th)\,\,.
\EN

The basic properties of form factors on kink states were discussed in 
Ref.\,\cite{DC}. The relations (\ref{fz1}), (\ref{fz2}) immediately lead to 
the equations
\bea
&& F^\Phi_{0i}(\th)=A_0(\th)F^\Phi_{0i}(-\th)+
      A_1(\th)\sum_{j\neq i}F^\Phi_{0j}(-\th)\,,\label{ffuni1} \\
&& F^\Phi_{i0}(\th)=B_0(\th)F^\Phi_{i0}(-\th)\,\,.
\eea
The two matrix elements $F^\Phi_{0j}(\th)$ and $F^\Phi_{j0}(\th)$ are 
related by crossing
\EQ
F^\Phi_{0j}(\th+2i\pi)=F^\Phi_{j0}(-\th)\,,
\EN
and in general have a simple pole at $\th=i\pi$ with residue
\EQ
-i\mbox{Res}_{\th=i\pi}F^\Phi_{0j}(\th)=i\mbox{Res}_{\th=i\pi}F^\Phi_{j0}(\th)=
\langle\Omega_0|\Phi|\Omega_0\rangle-\langle\Omega_j|\Phi|\Omega_j\rangle
\,\,.
\EN
The vacuum expectation values of $\Theta(x)$ give the vacuum energy densities 
and then are identical for the $q+1$ degenerate vacua. For the other two 
operators, colour symmetry leads to
\bea
&& \langle\Omega_0|\var|\Omega_0\rangle=U_0\,,\hspace{1cm}
   \langle\Omega_i|\var|\Omega_i\rangle=U_1\,,\\
&& \langle\Omega_0|\sigma_j|\Omega_0\rangle=0\,,\hspace{1cm}
   \langle\Omega_i|\sigma_j|\Omega_i\rangle=\frac{q\delta_{ij}-1}{q-1}\,M\,\,.
\eea
Here, $U_0-U_1$ gives the discontinuity in the internal energy across the first
order transition surface, and $M$ is the spontaneous magnetisation in the 
ordered phases. As a last necessary condition, the two-kink form factors are 
subject to the asymptotic bound \cite{immf}
\EQ
\lim_{\th\rightarrow +\infty}F^\Phi_{0i}(\th)\leq\mbox{constant}\,\,
e^{X_\Phi\th/2}\,\,,
\EN
with an analogous relation for $F^\Phi_{i0}(\th)$.

The above requirements uniquely determine the solutions\footnote{The form
factor of $\Theta(x)$ is normalised by the condition $F^\Theta_\pm(i\pi)=
2\pi m^2$, $m$ being the mass of the kinks.}
\bea
&& F^\Theta_{\pm}(\th)=-\frac{4i\pi}{p}m^2e^{\pm\frac{\gamma}{2}(\pi+i\th)}\,
\frac{\cosh\frac{\th}{2}}{\sinh\frac{1}{p}(\th-i\pi)}\,F_0(\th)\,\,,\\
&& F^\var_{\pm}(\th)=\pm i(U_1-U_0)\,
\frac{e^{\pm\frac{\gamma}{2}(\pi+i\th)}}{p\sinh\frac{1}{p}(\th-i\pi)}\,
F_0(\th)\,\,,\\
&& F^\sigma_{\pm}(\th)=\mp\frac{M}{2\Upsilon_+(i\pi)}\,
\frac{e^{\pm\frac{\gamma}{2}(\pi+i\th)}}{\cosh\frac{\th}{2}}\,
\Upsilon_\pm(\th)F_0(\th)\,\,,
\eea
with
\EQ
F_0(\th)=-i\sinh\frac{\th}{2}\,\exp\left\{\int_0^\infty\frac{dx}{x}\,
\frac{\sinh(1-p)\frac{x}{2}}{\sinh\frac{px}{2}\cosh\frac{x}{2}}\,
\frac{\sin^2(i\pi-\th)\frac{x}{2\pi}}{\sinh x}\right\}\,\,,
\EN
\EQ
\Upsilon_+(\th)=\exp\left\{2\int_0^\infty\frac{dx}{x}\,
\frac{\sinh(\frac{p}{2}-1)x}{\sinh\frac{px}{2}}\,
\frac{\sin^2(2i\pi-\th)\frac{x}{2\pi}}{\sinh 2x}\right\}\,\,,
\EN
\EQ
\Upsilon_-(\th)=\Upsilon_+(\th+2i\pi)\,\,.
\EN
The functions $F_0(\th)$ and $\Upsilon_+(\th)$ are solutions of the 
equations
\EQ
F_0(\th)=S_0(\th)F_0(-\th)\,,\hspace{1cm}F_0(\th+2i\pi)=F_0(-\th)\,,
\EN
\EQ
\Upsilon_+(\th)=\frac{\sinh\frac{1}{p}(i\pi+\th)}{\sinh\frac{1}{p}(i\pi-\th)}\,
\Upsilon_+(-\th)\,,\hspace{1cm}\Upsilon_+(\th+4i\pi)=\Upsilon_+(-\th)\,,
\EN
and behave as 
\EQ
F_0(\th)\sim\exp\left[\left(1+\frac{1}{p}\right)\,\frac{\th}{4}\right]\,\,,
\EN
\EQ
\Upsilon_+(\th)\sim\exp\left[\left(1-\frac{2}{p}\right)\,\frac{\th}{4}\right]
\,\,,
\EN
when $\th\rightarrow +\infty$.

\resection{Correlation functions}

In the $S$-matrix approach, correlation functions are expressed as spectral 
series over complete sets of intermediate asymptotic states. The two-kink
form factors we determined above are what is needed to compute the first
non-trivial term of these series for the operators $\Theta$, $\var$ and 
$\sigma_j$. Depending on the phase in which correlations are computed, for the 
two-point case we have
\bea
\langle\Omega_0|\Phi_1(x)\Phi_2(0)|\Omega_0\rangle_c=\sum_{i=1}^q
\int_{\th_1>\th_2}\frac{d\th_1}{2\pi}\frac{d\th_2}{2\pi}
F^{\Phi_1}_{0i}(\th_1-\th_2)F^{\Phi_2}_{0i}(\th_2-\th_1)\,e^{-|x|E_2}+
O(e^{-4m|x|}),\label{approx1}\\
\langle\Omega_i|\Phi_1(x)\Phi_2(0)|\Omega_i\rangle_c=
\int_{\th_1>\th_2}\frac{d\th_1}{2\pi}\frac{d\th_2}{2\pi}
F^{\Phi_1}_{i0}(\th_1-\th_2)F^{\Phi_2}_{i0}(\th_2-\th_1)\,e^{-|x|E_2}+
O(e^{-4m|x|}),\label{approx2}
\eea
where $\langle\cdots\rangle_c$ denotes connected correlators and 
$E_2=m(\cosh\th_1+\cosh\th_2)$ is the energy of the two-kink asymptotic state. 
The spectral series over form factors are 
known to converge quite rapidly (see \cite{DC} and references therein). A 
very effective way to test this property is to use the exact sum rules
\cite{cth,DSC}
\bea
&& C=\frac{3}{4\pi}\int d^2x\,|x|^2\langle \Omega_\alpha|\Theta(x)\Theta(0)|
\Omega_\alpha\rangle_c\,,\\
&& X_\Phi=-\frac{1}{2\pi\langle\Omega_\alpha|\Phi|\Omega_\alpha\rangle}\,
\int d^2x\,\langle \Omega_\alpha|\Theta(x)\Phi(0)|\Omega_\alpha\rangle_c\,,
\eea
($\alpha=0,1,\ldots,q$) to recover the conformal data (central charge and 
scaling dimensions) from the off-critical theory. 
The results obtained for $C$, $X_\var$ and $X_\sigma$ by plugging into these 
sum rules the two-kink approximations (\ref{approx1}), (\ref{approx2}) for the
correlators are shown in Figs.\,4,5,6 and compared with the exact formulas;
the numerical values corresponding to integer $q$ are given in the Table.
The growth of the scaling dimensions of the considered operators
with increasing $q$
leads to more severe ultraviolet singularities of the exact correlators which 
account for the decreasing accuracy of the two-kink approximation (which 
of course cannot reproduce the singularities). 

At $q=1$, the two-kink computation gives exact results for $C$ and $X_\var$
but not for $X_\sigma$. This can be understood as follows. With an obvious
simplification of notation, the $2(n+1)$-kink contribution to the correlator
$\langle\Omega_i|\Theta(x)\Phi(0)|\Omega_i\rangle$ can be written as
\EQ
\sum_{i_1,\ldots,i_n=1}^q\langle\Omega_i|\Theta|K_{i0}K_{0i_1}K_{i_10}\ldots
K_{0i_n}K_{i_n0}K_{0i}\rangle\langle K_{i0}K_{0i_1}K_{i_10}
\ldots K_{0i_n}K_{i_n0}K_{0i}|\Phi|
\Omega_i\rangle\,,
\label{unbroken}
\EN
or, for our present purpose, as
\bea
\langle\Omega_i|\Theta|K_{i0}K_{0i}\ldots K_{i0}K_{0i}\rangle
\langle K_{i0}K_{0i}\ldots K_{i0}K_{0i}|\Phi|\Omega_i\rangle+\nonumber\\
\sum{}'_{i_1,\ldots,i_n}\langle\Omega_i|\Theta|K_{i0}K_{0i_1}\ldots
K_{i_n0}K_{0i}\rangle\langle K_{i0}K_{0i_1}\ldots K_{i_n0}K_{0i}|\Phi|
\Omega_i\rangle\,,
\label{broken}
\eea
with the prime denoting the omission in the sum of the term with
$i_1=i_2=\ldots=i_n=i$.  
The first term in (\ref{broken}) involves only two vacua and at $q=1$ is 
computed on the physical excitations of the thermal Ising model, which behave 
as free fermions (scattering amplitude equal $-1$). As a consequence, the 
$2(n+1)$-kink form factors
$\langle\Omega_i|\Theta|K_{i0}K_{0i}\ldots K_{i0}K_{0i}\rangle$
vanish at $q=1$ for any $n>0$. Consider now the primed sum in (\ref{broken}).
It involves $q^n-1$ terms wich, if $\Phi$ is invariant under colour
permutations, are all identical and finite. Therefore, we conclude that, 
at $q=1$, (\ref{unbroken}) is identically zero for any $n>0$ whenever $\Phi$ 
is a colour singlet operator (e.g. $\Phi=\Theta,\var$). The situation is more 
subtle when $\Phi=\sigma_i$. In fact, although the number of terms in the 
primed sum vanishes as $q\rightarrow 1$, now the sum contains contributions 
which diverge in the same limit\footnote{An example of this mechanism is 
obtained specialising Eq.\,(\ref{ffuni1}) to the case $\Phi=\sigma_i$
and recalling Eq.\,(\ref{constraint}).}. In this way, the correlator
$\langle\Omega_i|\Theta(x)\sigma_i(0)|\Omega_i\rangle_{q=1}$ receives a finite 
contribution even from excitations which are unphysical at $q=1$. This is 
not that surprising if one considers that the operator $\sigma_i$ itself
is unphysical at $q=1$ (there are no Potts spin degrees of freedom) and only 
makes sense as an analytic continuation.

We conclude with two remarks concerning correlation lengths and interfacial
tensions. The ``true'' correlation lengths $\xi_0$, $\xi_i$ in the 
disordered and ordered phases are defined through the large distance decay
of the spin-spin correlation functions
\EQ
\langle\Omega_\alpha|\sigma_i(x)\sigma_i(0)|\Omega_\alpha\rangle_c\sim
e^{-|x|/\xi_\alpha}\,\,,\hspace{1cm}\alpha=0,i\,\,.
\EN
Since this asymptotic behaviour is given by Eqs.\,(\ref{approx1}), 
(\ref{approx2}), one immediately concludes that
\EQ
\xi_0=\xi_i=\frac{1}{2m}\,\,.
\label{true}
\EN
In two dimensions the interfacial tension between two coexisting phases 
coincides with the mass
of the excitation which interpolates between them. Then, denoting $\sigma_{i0}$
the ordered-disordered tension and $\sigma_{ij}$ ($i\neq j$) the 
ordered-ordered tension, we have
\EQ
\sigma_{i0}=\frac{\sigma_{ij}}{2}=m\,\,.
\label{tension}
\EN

\resection{Conclusion}

In this paper we applied the $S$-matrix program to the two-dimensional dilute 
$q$-state Potts model on its first order phase transition surface for 
$q\leq 4$. We have shown how integrability allows an exact scattering 
description of the scaling limit and how this leads to the computation of 
correlation functions for the interesting operators through the form factor
approach. The sum rules for the central charge and the scaling dimensions
have been used to show the remarkable accuracy provided by the two-kink
approximation for the correlators. Of course, in the same approximation, the
correlation functions computed here can be used to evaluate other universal
quantities which are characteristic of the off-critical model and are not
known from conformal field theory, for example the universal combinations
of critical amplitudes. For the pure model, this was done in Ref.\,\cite{DC}.

We have shown how the results (\ref{true}) and (\ref{tension}) for the 
correlation lengths in the different coexisting phases and the interfacial
tensions are simple consequences of the basic structure of the scattering 
theory associated to the statistical model. Although these results were 
obtained for $q\leq 4$, there are reasons to believe that they hold true for 
$q>4$. As a matter of fact, in the pure system at $q>4$, Eq.\,(\ref{tension})
is known to hold \cite{BJ}, and Eq.\,(\ref{true}) received strong
support form Monte Carlo simulations \cite{JK}.

\vspace{1cm}
\noindent
{\bf Acknowledgements.} I am grateful to John Cardy for interesting 
discussions.



\newpage

\begin{center}

\begin{tabular}{|c||c|c|c|c|}\hline
$ q $ & $1$  &  $2$        & $3$       & $4$      \\ \hline
$ C $ & $1/2$ & $ 7/10 $ & $ 6/7 $  &  $ 1 $  \\
$   $ & $1/2$ & $ 0.699$ & $ 0.853$ & $0.987$\\  \hline
$X_\var$&$1/8$& $ 1/5 $  &  $2/7$   & $1/2$ \\
$   $ & $ 1/8 $ &$0.202 $ & $0.295$ & $0.560$ \\ \hline
$X_\sigma$&$5/96$& $3/40$ & $2/21$  & $1/8$ \\
$   $ & $0.0514$ & $0.0734$ & $0.0919$ & $0.101$ \\ \hline
\end{tabular}
\end{center}
\vspace{1.5cm}
{\bf Table.} Central charge and scaling dimensions of the thermal and spin 
operators in the
tricritical $q$-state Potts model. The results of the two-kink approximation 
are shown below the exact values.

\vspace{3cm}

\begin{figure}
\centerline{
\psfig{figure=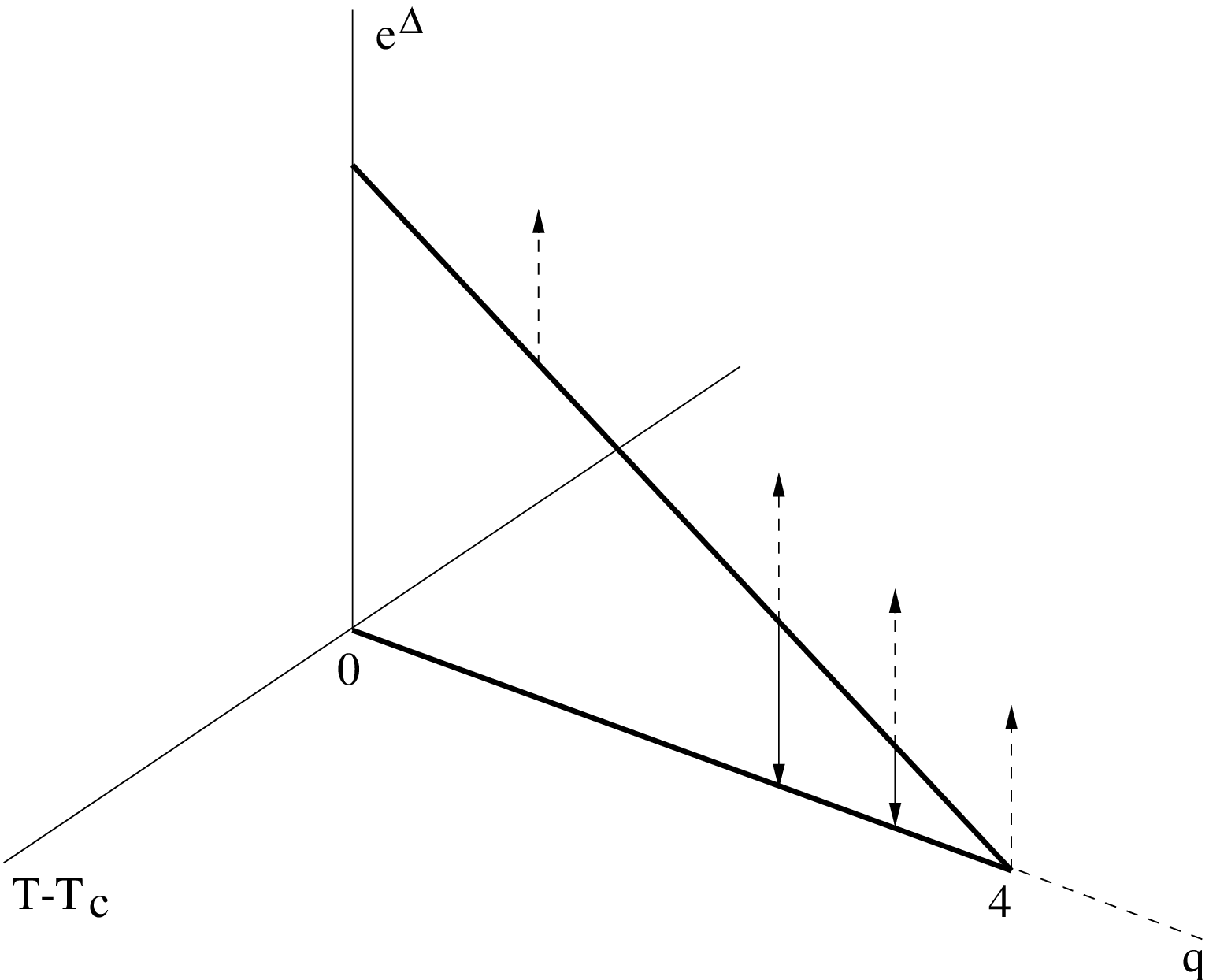}}
\vskip 1.5cm
{\bf Figure 1.} Phase diagram of the dilute $q$-state Potts model in the
space of temperature, vacancy density and number of states.
\end{figure}

\newpage
\begin{figure}
\centerline{
\psfig{figure=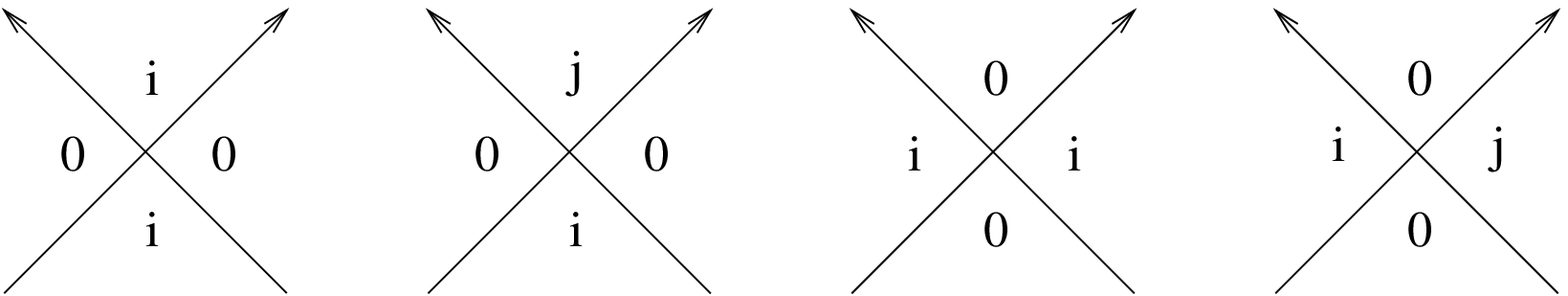}}
\vskip 1.5cm
{\bf Figure 2.} The two-kink scattering amplitudes $A_0$, $A_1$, $B_0$, 
$B_1$ ($i\neq j$).
\end{figure}


\begin{figure}
\centerline{
\psfig{figure=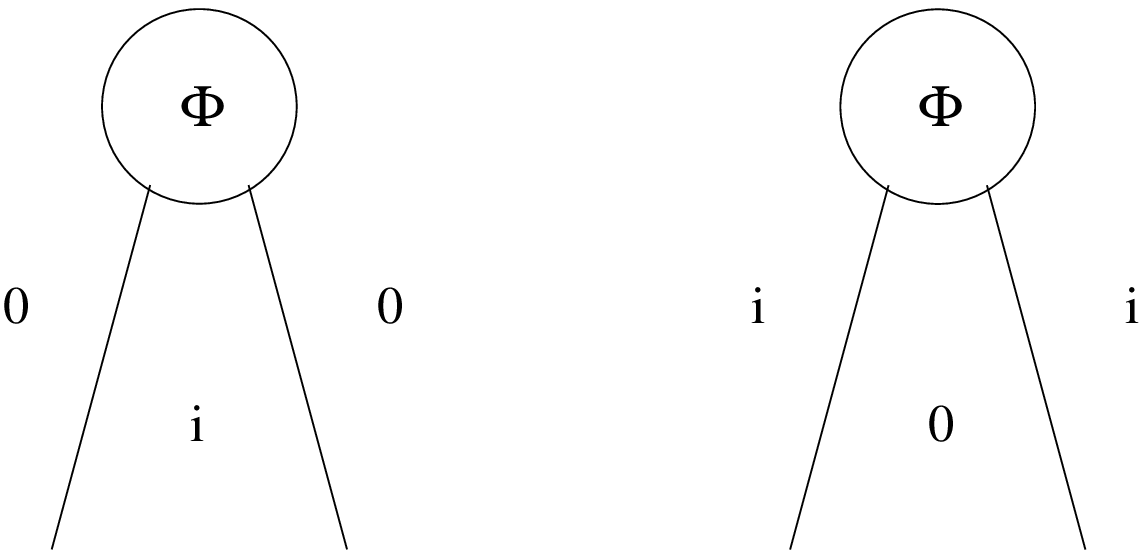}}
\vskip 1.5cm
{\bf Figure 3.} The two-kink form factors $F^\Phi_{0i}$ and $F^\Phi_{i0}$
of a colour singlet operator $\Phi$.
\end{figure}

\newpage
\begin{figure}
\null\vskip -4cm 
\centerline{
\psfig{figure=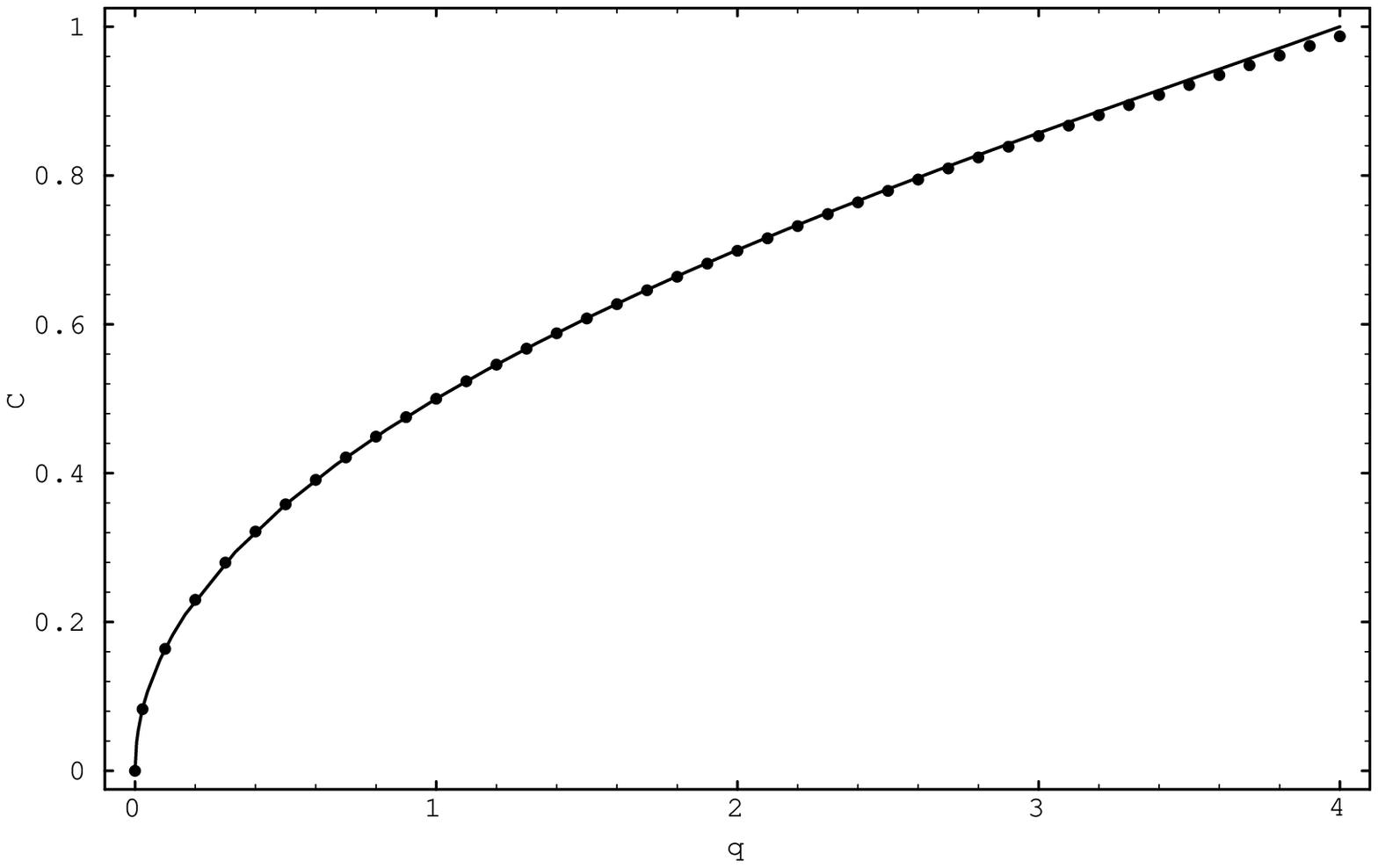}}
\vskip -5cm
{\bf Figure 4.} Central charge of the tricritical $q$-state Potts model. 
Exact formula (\ref{c}) (continous line) and two-kink approximation (dots).
\end{figure}

\newpage
\begin{figure}
\null\vskip -4cm 
\centerline{
\psfig{figure=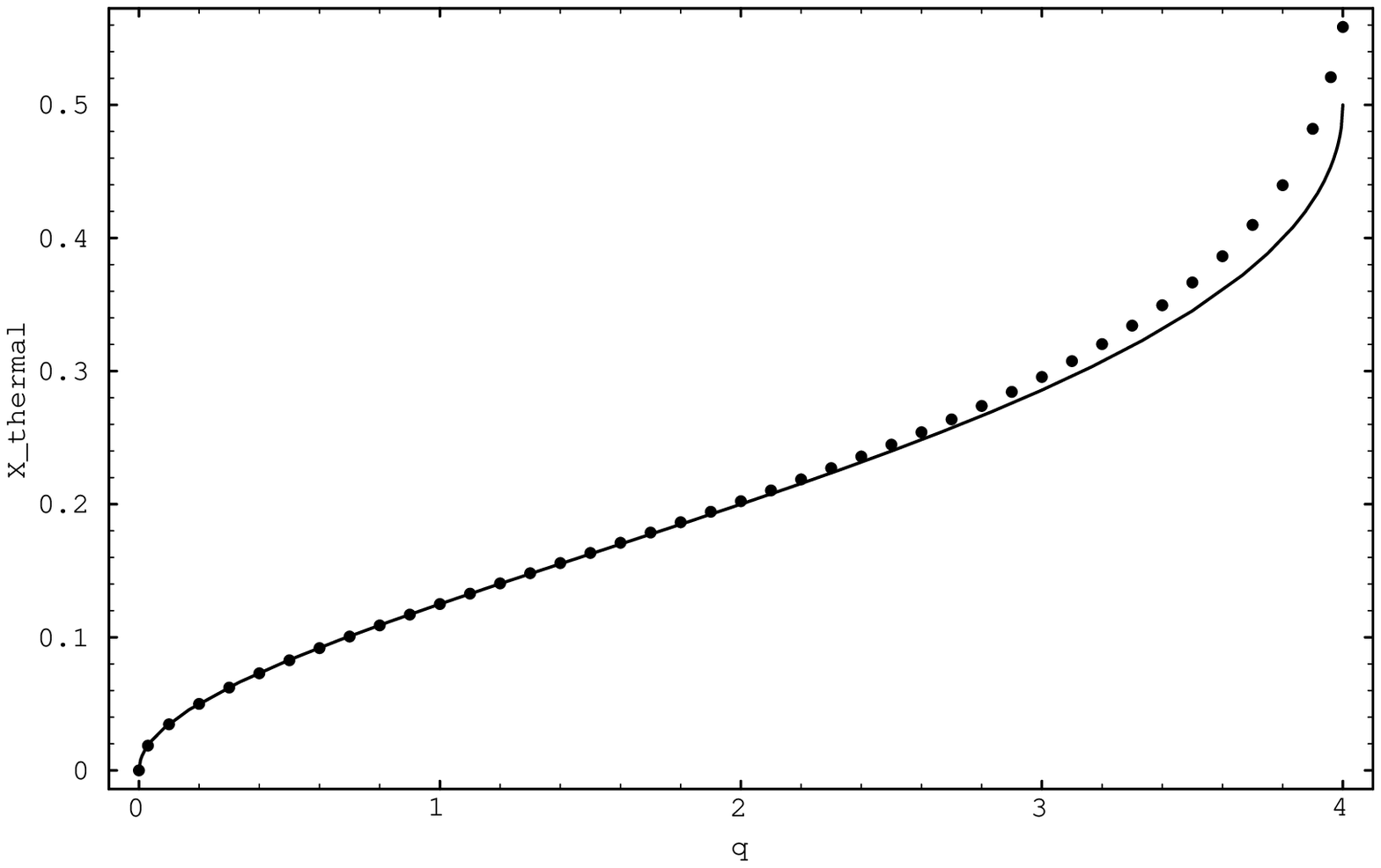}}
\vskip -5cm
{\bf Figure 5.} Scaling dimension of the thermal operator in the tricritical 
$q$-state Potts model. Exact formula (\ref{x12}) (continous line) and two-kink
approximation (dots).
\end{figure}

\newpage
\begin{figure}
\null\vskip -4cm 
\centerline{
\psfig{figure=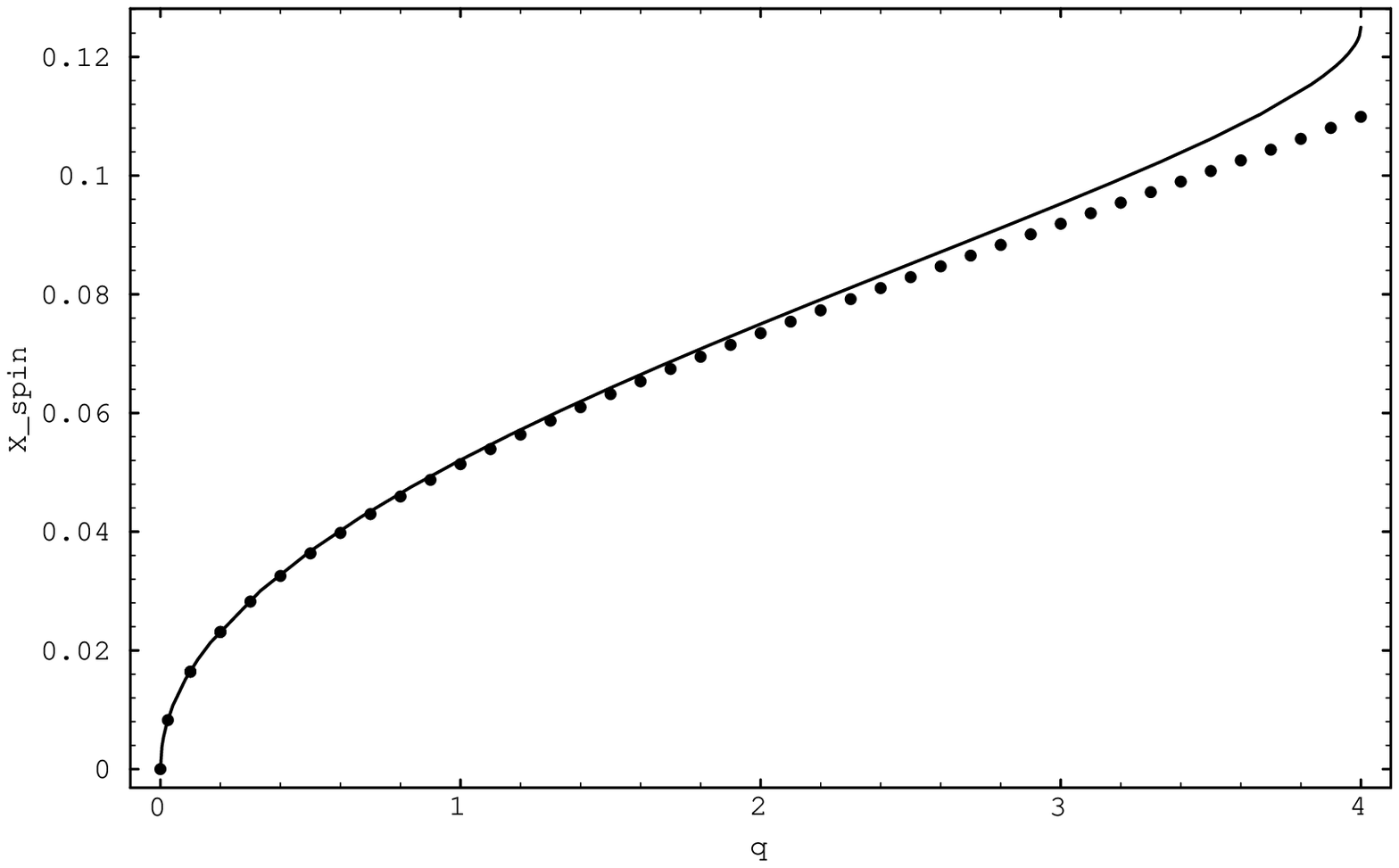}}
\vskip -5cm
{\bf Figure 6.} Scaling dimension of the spin operators in the tricritical 
$q$-state Potts model. Exact formula (\ref{xsig}) (continous line) and two-kink
approximation (dots). 
\end{figure}

\end{document}